\documentclass[aps,prl,twocolumn,groupedaddress,10pt]{revtex4}
\usepackage{graphicx}
\usepackage[english]{babel}

\begin{document}

\title{\bf The unintended influence of control systems on edge-plasma 
 \\ transport and stability in the Joint European Torus}
 
\author{A. J. Webster$^{1}$}
\author{J. Morris$^{1}$}
\author{T. N. Todd$^{1}$}
\author{S. Brezinsek$^{2}$}
\author{JET EFDA Contributors\footnote{See the Appendix of
    F. Romanelli et al., Proceedings of the 24th IAEA Fusion Energy
    Conference 2012, San Diego, US.}}

\affiliation{JET-EFDA, Culham Science Centre, Abingdon, OX14 3DB, UK}
\affiliation{$^1$ CCFE, Culham Science Centre, Abingdon, OX14 3DB, UK} 
\affiliation{$^2$ Institut f\"ur Energie- und Klimaforschung -
  Plasmaphysik, Forschungszentrum J\"ulich, 52425 J\"ulich, Germany.} 


\date{\today}

\begin{abstract}
A unique experiment in the Joint European Torus (JET) consecutively
produced 120 almost identical plasma pulses, providing two orders of
magnitude more data than is usually available.    
This allows the statistical detection of previously unobservable
phenomena such as a sequence of resonant-like waiting times between
edge-localised instabilities (ELMs).   
Here we investigate the causes of this phenomenon. 
By synchronising data to the 1000s of ELM times and averaging the
results, random errors are reduced by a factor of 50, allowing
unprecedentedly detailed behaviour to be described.  
A clear link can then be observed between plasma confinement, ELM
occurrence, vertical plasma oscillations, and an otherwise
unobservable oscillation in a control coil current that is not usually
associated with ELM occurrence.  
The results suggest a strong and unanticipated edge-plasma
dependence on control system behaviour. 

\end{abstract}

\maketitle

The successful operation of a large magnetically confined fusion
plasma experiment in a tokamak such as ITER \cite{ITER}, will require
the successful mitigation of potentially damaging edge-localised
plasma instabilities (ELMs) \cite{Zohm,Kamiya}.   
Unless they are controlled or avoided, an economically viable fusion 
power plant seems an unrealistic prospect.  
Here we describe observations that point to an unexpected new method
for modifying the edge plasma stability and density. 
The origin of this discovery is the analysis of data from over 120
consecutively created 2T, 2MA, high confinement (H-mode) plasmas in
the Joint European Torus (JET). 
These were designed to study the
movement of eroded plasma 
facing material within JET's vacuum vessel after the installation of
JET's new ITER-like wall \cite{ILW}, full details are in \cite{Brezinsek}. 
The data set is unprecedentedly large, with each plasma having about 6
seconds of H-mode with type I ELMs. 
Combining data from the exceptionally steady-state final 2 seconds of
H-mode, provides 240 seconds of plasma with $\sim$10000 ELMs. 

\begin{figure}[htbp!]
\begin{center}
\includegraphics[width=8cm,height=5.cm]{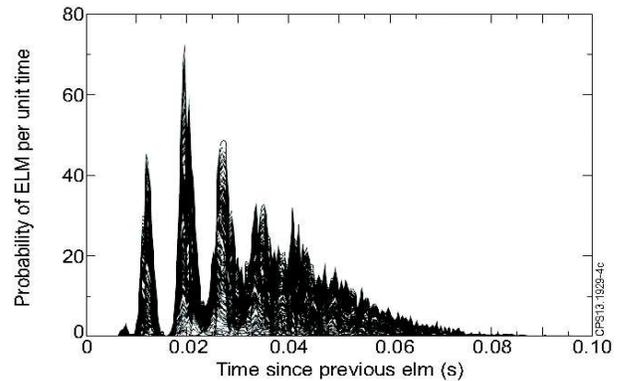}
\vspace{.5cm}
\caption{\label{Tree} 
The ELM waiting-time data from 120 almost identical JET plasmas is
combined to form a single pdf for the waiting times between ELMs. 
Each line corresponds to data from an additional pulse, that are added 
together to form the pdf. 
Reproduced from Ref. \cite{resonances}.}
\end{center}
\end{figure}

\begin{figure}[htbp!]
\begin{center}
\includegraphics[width={8.5cm},height=8cm]{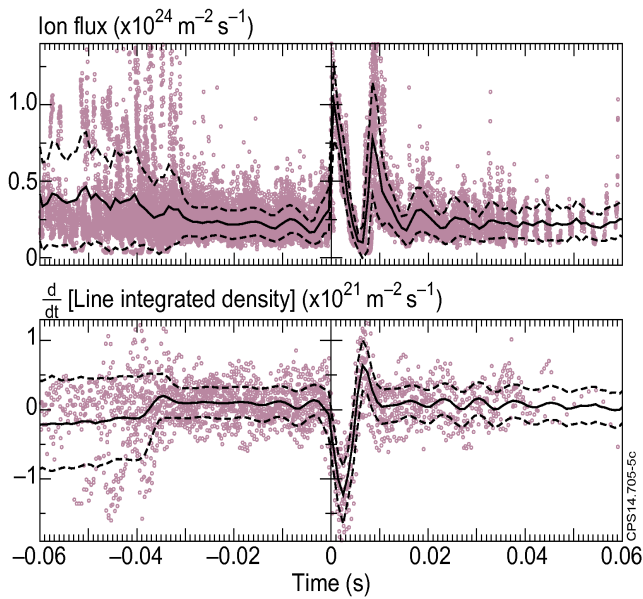}
\caption{\label{ionflux} 
Top - Langmuir probe measurements of ion flux to the inner
divertor ($\times 10^{24}$ m$^{-2}$s$^{-1}$) versus time (s), 
synchronised to the ELM times at $t=0$. 
Bottom - the rate of change of line-integrated density measured
through the plasma's mid-plane (particles m$^{-2}$ s$^{-1}$) versus
time (s), synchronised to the ELM times at $t=0$. 
Thick black lines are averages, dashed lines indicate 
standard deviations, and circles are typical measurements 
(pulse 83794).  
Prior to ELMs there are 8ms-period oscillations, with increased (or
reduced) ion fluxes that coincide with reducing (or increasing) plasma
density, similar to 8ms-period plasma-position and control-current 
oscillations discussed later. 
The post-ELM signal is difficult to interpret due to large post-ELM
plasma movements, strong control coil responses, and non-linear
affects such as impurity influxes. 
}
\end{center}
\end{figure}

\begin{figure*}[htbp!]
\begin{center}
\includegraphics[height=6.cm,width=\columnwidth]{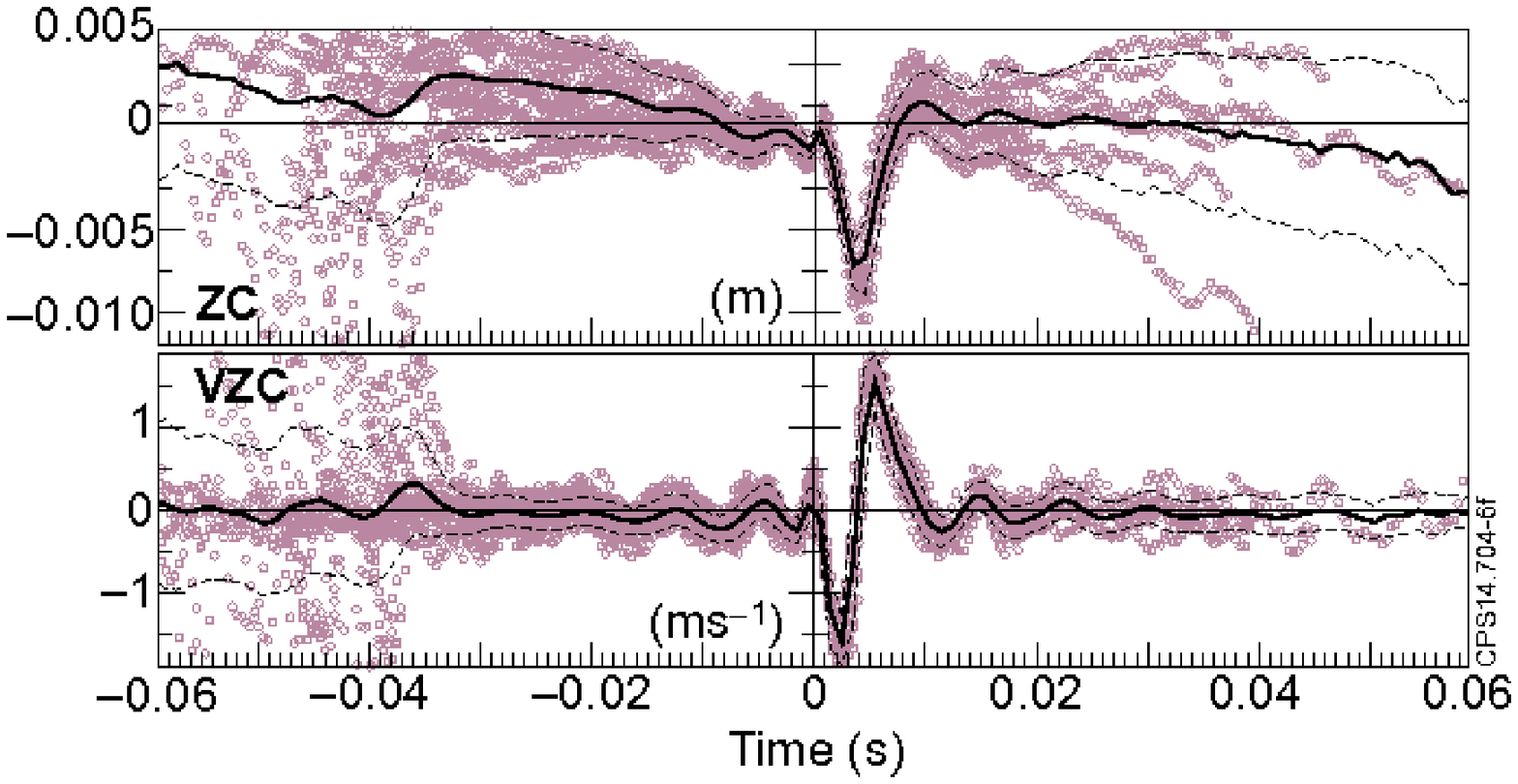}
\includegraphics[height=6cm,width=\columnwidth]{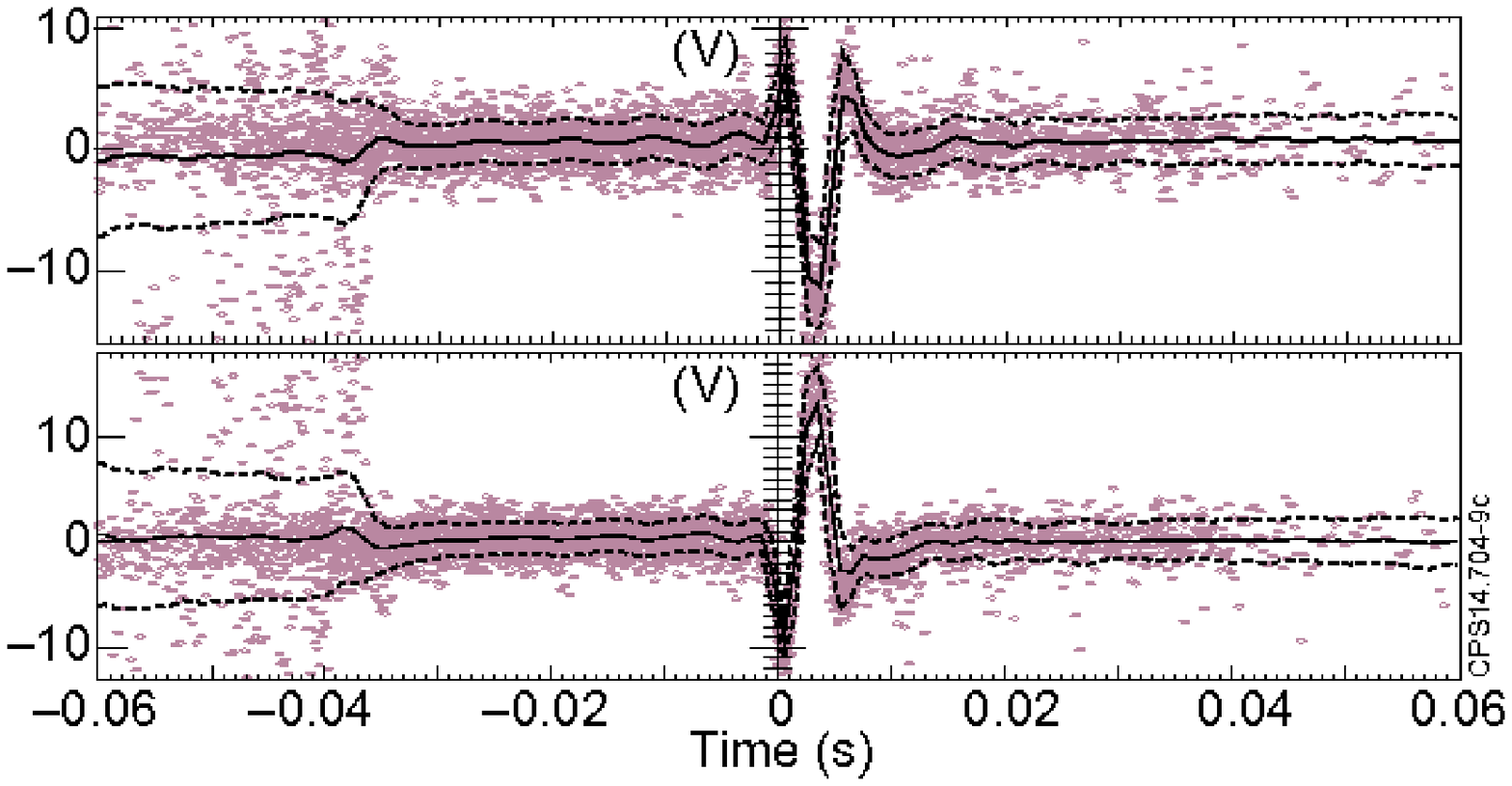}
\caption{\label{EMFs} 
  Top and bottom left are the vertical position (ZC) and velocity
  (VZC) of the plasma current's centre, as calculated by EFIT
  \cite{EFIT,EFIT2}. 
  Top and bottom right are the EMFs measured by toroidal loops above
  and below the plasma respectively. 
  The measured EMFs are $\pi$ out of phase, consistent with a vertical
  oscillation, and have a phase and amplitude consistent with the
  vertical oscillations calculated by EFIT. 
}
\end{center}
\end{figure*}

\begin{figure*}[htbp!]
\begin{center}
\includegraphics[height=6.5cm,width=\columnwidth]{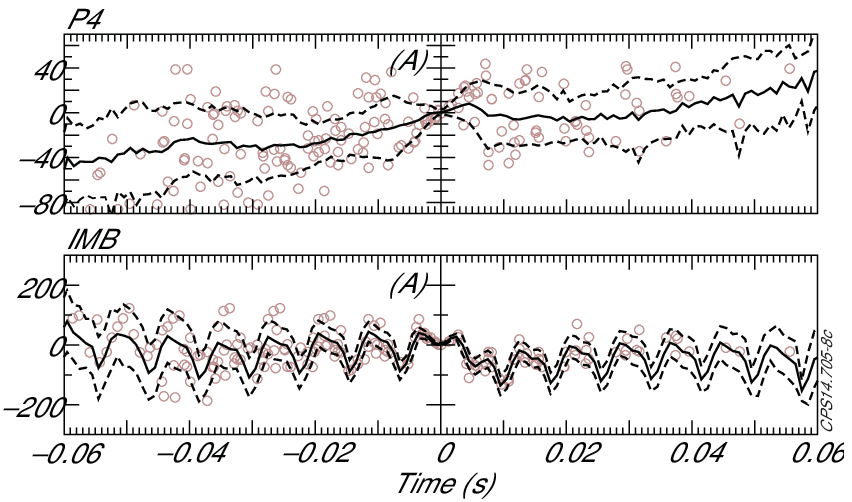}
\includegraphics[height=6.5cm,width=\columnwidth]{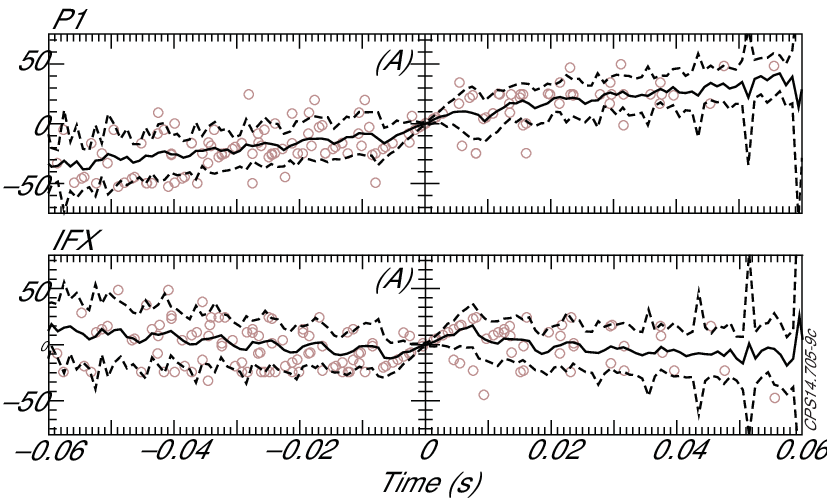}
\caption{\label{PCS} 
  Changes in control system currents (Amps) versus time (s), relative
  to those at ELM time $t=0$. 
  Clockwise from top left, circuits: P4, P1, PFX, and IMB. 
  Circles are a typical pulse (83794), pulse-set averages and
  standard deviations are the solid and dashed lines respectively.  
  The 8ms period oscillations in IMB are very clear, and its standard
  deviation is comparatively very small.   
  Note that these are oscillations in current, not voltage, and
  correspond to magnetic perturbations of the plasma. 
}
\end{center}
\end{figure*}

The large quantities of data produced the most finely resolved
probability density function (pdf) for the waiting times between ELMs
to date. 
In contrast to expectations from previous work \cite{WebsterPRL}, the
pdf was found to have a sequence of maxima and minima separated by 8ms
intervals, with the first maxima 12ms after the previous ELM 
(figure \ref{Tree}) \cite{resonances}. 
Here we explore the cause of this phenomenon. 
Key to our analysis is the unprecedentedly large amount of ELM data; 
each ELM is in principle, statistically equivalent. 
By synchronising data to the ELM times, and averaging over e.g. 3000 
ELMs, the central limit theorem ensures that random errors will be
reduced by a factor of $1/\sqrt{3000}\simeq 0.02$. 
This allows detail to be observed that would not otherwise be
possible, and enables the analysis presented here. 
The Berylium II (527nm) radiation measured at the inner divertor in
0.1ms intervals is used to identify ELM-times, as in
\cite{WebsterPRL}. 
When synchronising the data to ELM times, we deliberately exclude data
for which the previous ELMs occurs within 40 ms. 
This is to exclude any large post-ELM responses from being included in
the pre-ELM signal, but reduces the number of ELMs to $\sim$3000.  
The resulting plots (figures \ref{ionflux}, \ref{EMFs}, and \ref{PCS})
include the average and its standard deviation, plus values from a
typical JET pulse in the set (83794).   
The Supplementary Material \cite{SM} lists the 
pulses that we consider. 

Figure \ref{ionflux} plots Langmuir probe measurements of Deuterium
ion fluxes to JET's inner divertor, and the rate of change of 
line-integrated plasma density measured along a chord through the
plasma's mid-plane. 
Because the chord is through the mid-plane, the line-integral is
insensitive to small vertical plasma displacements, and should solely
measure changes in plasma density. 
Both plots show 8ms-period oscillations in the ion flux and
rate of change of plasma density, with increased ion fluxes
coinciding with decreasing density, and vice versa. 
Line-integrated edge-density measurements are similar. 

Figure \ref{EMFs} shows: the vertical position and velocity of the
plasma current's centre as calculated by EFIT \cite{EFIT,EFIT2}, and
electromotive forces (EMFs) measured by toroidal ``flux'' loops that
are vertically above and below the plasma. 
There are clear 8ms-period pre-ELM oscillations in EFIT's calculated
plasma motion, that appear to be confirmed by EMFs that are measured
by the two flux loops.  
The measured EMFs are $\pi$ out of phase with each other, 
consistent with a vertical plasma oscillation, and have a phase and
amplitude that is consistent with the vertical velocity calculated
by EFIT. 
It is well known that vertical plasma displacements can modify plasma
stability \cite{Liang}, and the oscillation's period is the
same as for enhanced (or reduced) ELM occurrence and ion losses, so it
is possible that they are triggered by the vertical oscillations. 
The maxima in ELM occurrence and ion fluxes are when the plasma is
moving rapidly towards its furthest downwards displacement, 
and the minima in ELM occurrence and ion fluxes have the plasma near 
its uppermost position.   
The correlation between EMFs, plasma motion, and ELMs, 
explains the phase relationship observed between ELM occurrence and 
measured flux loop voltages in JET's divertor in \cite{Chapman}. 

Previous analysis \cite{resonances} indicated that the oscillations
observed in figures \ref{ionflux} and \ref{EMFs} must result from
either a plasma phenomena or a real time control system, but found no
evidence for the vertical control system being responsible.  
JET also has a real-time shape control system that alters 9 coil
circuits: 4 divertor coils, and circuits named as: P1, P4, SHA, IMB,
and PFX (see \cite{Sartori} for details). 
The divertor coils are used to control the plasma strike point
positions, P1 controls the plasma current, P4 controls the outer gap
(ROG), SHA controls the plasma's triangularity and elongation, IMB
controls the top gap (TOG), and PFX controls the inner 
gap (RIG); although as indicated in figure 2 of \cite{Sartori},
each circuit modifies the plasma's shape and position in a variety of
ways.  
The shape controller can modify these currents every 2ms, 
and they are recorded every 8ms, which is too infrequent for an 8ms
period oscillation to be observable in time traces of the currents.  
Consequently the 240s of data and 3000 ELMs become essential. 
Figure \ref{PCS} plots the measured currents after synchronising them
to the ELM times and offsetting them by their value at the ELM 
($t=0$), estimated by linearly interpolating between adjacent data values. 
The result is surprisingly successful, with comparatively small error
bars in many of the plots. 

\begin{figure}[htbp!]
\begin{center}
 \includegraphics[height=3.5cm,width=7cm]{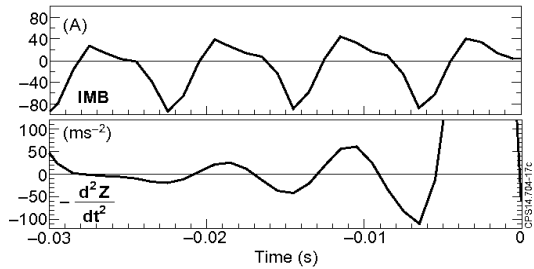}
\caption{\label{acc}
  Top figure, the average IMB current prior to ELMs, offset to zero at
  the ELM time. 
  Bottom figure,
  the average plasma acceleration downwards prior to ELMs, as measured
  by the accelleration at the plasma's X-point ($-d^2 Z /dt^2$).
  When the IMB oscillation is positive then there is a positive
  acceleration downwards,  when the oscillation is negative the plasma 
  accelerates upwards. 
}
\end{center}
\end{figure}

The divertor coil and SHA currents are comparatively unremarkable and
are not shown here.  
As are the P4 currents, that show a gradual increase prior to ELMs
but no oscillations. 
In contrast, P1, PFX, and IMB, show clear oscillations with an 8ms
timescale prior to ELMs.  
It is worth emphasising that these plots represent measured currents
in the circuits (not voltages), with associated magnetic fields that
directly displace the plasma's shape and position.  
For these plasmas the control system settings use P1 to keep the
plasma current constant, the PFX currents are not used to control RIG
but are requested to be kept a constant fraction of the plasma's
current (that is approximately constant), and use IMB to control TOG. 
The currents in IMB in particular show clear oscillations.
The error bars on these average IMB currents are very small,
indicating that this 150A current oscillation has a very consistent
8ms period. 
This seems likely to be the cause of the oscillatory
observations described here and elsewhere \cite{resonances}. 
The ELMs occur 4ms after a maximum in IMB, which is 12ms from the
previous maximum. 
The maxima are at 12, 20, 28, 36, 44, and 52 ms before and
after an ELM; identical to the timings of the maxima in figure 2 of
\cite{resonances}. 
An increase in IMB current would be expected to pull the plasma down
and outwards \cite{Sartori}, and a decrease in IMB to push the plasma
up and inwards.  
Figure \ref{acc} is consistent with this, finding a downwards plasma  
acceleration when the oscillation in IMB current is +ve, and an 
upwards acceleration when the oscillation in IMB current is -ve.  
The ELMs (and enhanced ion fluxes), occur as IMB pulls the plasma
outward and downwards.  
The size and coherence of the oscillations in IMB make a 
persuasive case for IMB being the cause of the oscillatory phenomena
discussed in this paper.  
As does their exact coincidence with enhanced (or reduced) ion fluxes
and ELM occurrence, and the correlation between the sign of the
IMB current's oscillation and the direction of plasma acceleration. 
How the oscillations arise is unclear, but it is known that
oscillations can easily arise in a control system involving 9 
independent circuits and the plasma \cite{Sartori}. 
The oscillations are not universally present in JET plasmas, and
depend on plasma heating and fueling for example \cite{resonances},
and could involve a coupling between various independent circuits, the
plasma, and the plasma's motion.  
A rigorous analysis is likely to require both a faster recording of
control coil currents and an effective modelling of the plasma's
response to them. 
We emphasise that because the currents in these circuits were
only recorded every 8ms, it is impossible to get an indication that
the oscillations were present from the time series data alone - this
was only possible due to the large number of ELMs and plasma pulses in
the analysis. 
We also emphasise that the oscillations in IMB current could arise
from a complex process or sequence of processes involving one or more
of: plasma instabilities, turbulence, transport processes, material
erosion and recycling 
for example, 
but we think it simplest to regard the IMB 
current as the ``cause'' of the vertical oscillations and to search for
the cause or sequence of processes by which the IMB oscillations are 
produced. 

To summarise, an unprecedented number of 120 identical JET plasmas has
allowed the observation of a totally unexpected connection between a
seemingly benign shape control system, ELM occurrence, and edge transport. 
There are numerous implications. 
Firstly, these and similar effects could be common but unobserved, due
to insufficient data. 
This has clear implications for the analysis of experiments for ELM
pacing in particular \cite{Liang}, there could be synergies favouring
different ELM frequencies that need to be tested. 
Second, it suggests that a control system or a plasma-control system
interaction is causing vertical oscillations, and are either directly 
or indirectly modifying the plasma's edge-transport and stability. 
More significantly, it opens the possibility that entire classes of
ELMs could involve synergistic control system interactions - this will
be explored in greater detail elsewhere, and may require a     
modification of the paradigm whereby plasma stability and transport is 
studied independently of control system behaviour; and vice versa.      
Finally, the suggestion that IMB can modify edge transport rates leads
us to ask, could it modify edge densities sufficiently to  
avoid ELMs entirely? 
We hope these possibilities will be explored in the near future. 

{\bf Acknowledgements:} 
AJW thanks: 
Francesco Maviglia, David Keeling, Geoffrey Cunningham, Ian Chapman, 
Peter Lomas, Alan West, Paul Smith, Simon Hotchin, Andre Neto, 
Philip Andrew, Maximos Tsalas, Tobias Schlummer, and Ephrem Delabie, for 
help in various ways. 
This work was supported by EURATOM and carried out within the
framework of the European Fusion Development Agreement. 
For further information on the contents of this paper please contact 
publications-officer@jet.efda.org.
The views and opinions expressed herein do not necessarily reflect
those of the European Commission.  
The work was also part-funded by the RCUK Energy Programme under grant
EP/I501045. 

\end{document}